\documentclass[12pt,onecolumn]{IEEEtran}
\usepackage[T1]{fontenc}
\usepackage{amssymb}
\usepackage{amsmath}
\usepackage{amsthm}
\usepackage{amsfonts}
\usepackage{amsthm}
\usepackage{amsmath}
\usepackage{amsfonts}
\usepackage{amssymb}
\usepackage{bbm}
\usepackage{graphicx}
\usepackage{hyperref}
\usepackage{algorithm,algpseudocode}
\usepackage{mathtools}
\usepackage{dsfont}
\usepackage{verbatim}
\usepackage{subcaption}
\usepackage{float}
\usepackage{eucal}
\usepackage{balance}
\usepackage{bm}
\usepackage{amssymb}
\usepackage{amsmath}
\usepackage{amsthm}
\usepackage{amsfonts}
\usepackage{hyperref}
\usepackage{bbold}
\usepackage{mathtools}
\usepackage{dsfont}
\usepackage{verbatim}
\usepackage{subcaption}

\newtheorem{theorem}{Theorem}

\newtheorem{definition}[theorem]{Definition}

\title{Fast Beam Training for RIS-Assisted Uplink Communication}

\author{
\IEEEauthorblockN{Chandradeep Singh, Kamal Singh, K. H. Liu}
\vspace*{-0.65cm}
}

\begin{document}
\maketitle
\begin{abstract}
In this work, we propose a beam training codebook for Reconfigurable Intelligent Surface (RIS) assisted mmWave uplink communication. Beam training procedure is important to establish a reliable link between user node and Access point (AP). A codebook based training procedure reduces the search  time to obtain best possible phase shift by RIS controller to align incident beam at RIS in the direction of receiving node. We consider a semi passive RIS to assist RIS controller with a feedback of minimum overhead. It is shown that the procedure detects a mobile node with high probability in a short interval of time. Further we use the same codebook at user node to know the desired direction of communication via RIS.
\end{abstract}
\begin{IEEEkeywords}
Random Access , Beam Training, Beamforming, Codebook, Reconfigurable Intelligent Surface (RIS). 
\end{IEEEkeywords}

\section{Introduction}
\IEEEPARstart{T}{ransmission} at higher frequencies in mmWave communication is susceptible to severe pathloss. To compensate pathloss, highly directed antenna arrays are used at the transmitter and the receiver nodes. The directed beams used for the line of sight (LoS) communication at mmWave frequencies face severe attenuation due to blockage by the objects placed in between the LoS path. RIS usually deployed to provide virtual LoS communication between the AP and the mobile nodes to enhance the channel quality at higher frequencies. RIS is the key technology to achieve a smart reconfigurable wireless environment for next generation wireless networks \cite{Wu2020}. RIS smartly controls the reflection using large low cost energy efficient reflecting elements to enhance the signal strength or minimizing interference at the receiver. Random access (RA) is the key procedure in wireless networks that enables a mobile node to communicate with AP. If the mobile node is not in direct LoS with the AP then the RIS must assist the RA procedure. 

RA is well defined in literature available on 5G \cite{Arana2017}. Existing wireless technology has standards 802.11ad and 802.11ay for RA and beam training. These protocols also avoid beam collision in multi user case. AP provides channel access to different clients in the network based on polling. In \cite{Zhou2017}, enhanced RA procedure with beam training for high density networks is proposed which is compatible with 802.11ad and 802.11ay. In \cite{Shen2020}, multi-user association and beam training with flexible training slots is proposed to avoid collision. RA for millimeter wave non-orthogonal multiple access (NOMA) has been proposed in \cite{Ding2017}. In \cite{Hongbo2020}, hybrid grant RA scheme for maximal ratio combining (MRC) receiver and zero forcing (ZF) receiver is proposed for massive MIMO systems.

Aforementioned discussion shows the importance of beam training procedure in establishing a reliable link between AP and user node in mmWave communication. In \cite{Xiao2016}, a binary search tree based codebook is proposed for beam training. The codebook based on a binary search tree uses broad beams while starting a search and hence faces severe interference from multi paths at the receiver node as discussed in \cite{Hassanieh2018}. Binary tree based search is also not applicable to RIS assisted communication as it starts with deactivating the elements and in case of RIS assisted communication, received signal becomes very weak if we deactivate RIS elements. Hence a multi beam codebook is a possible solution for the RIS assisted communication. To the best of our knowledge, \cite{Zhang2020} is the first codebook proposed for RIS-assisted downlink communication. This work can not be applied to multi-user uplink communication as beams from different users collide at RIS-AP link during beam training at RIS. Due to destructive interference AP might receive low signal to noise ratio (SNR) even if multiple users are requesting for channel access. No significant work is available to detect and avoid beam collision for RIS assisted uplink communication. Hence we consider a simple case of single user for uplink communication and in future, this work can be extended to detect collision of beams from multiple users in uplink communication. Even the collision detection in multi user uplink communication also helps in beam training procedures for uplink communication.

\textbf{Notations:} Upper-case and lower-case boldface letters denote matrices and column vectors, respectively. Upper-case calligraphic letters (e.g.$\mathcal{F}$) denote discrete and finite sets. Superscripts $(.)^T,(.)^H\mbox{ and }(.)^{-1}$ stand for the transpose, Hermitian transpose, and matrix inversion operations, respectively. $\mathbb{C}^{a{\times}b}$ denotes the space of $a{\times}b$ complex-valued matrices. $\mbox{diag}\{a_1,\ldots,a_N\}$ denotes the diagonal matrix with elements $\{a_1,\ldots,a_N\}$. $\mathcal{CN}(0,\sigma^2)$ denotes a circularly symmetric complex Gaussian random variable with zero mean and variance $\sigma^2$.

The rest of the article is organized as follows: System model and problem formulation is discussed
in Section~\ref{sec:sys_model}. Our proposed beam training algorithm is described in Section~\ref{sec:algo}. 
Numerical evaluation of the proposed algorithm is provided in Section~\ref{sec:ne}.  
Finally, we conclude in Section~\ref{sec:con}.

\section{System Model and Problem Formulation}
\label{sec:sys_model}

\begin{figure}[h]
\vspace{1em}
\begin{centering}
\includegraphics[scale=0.5]{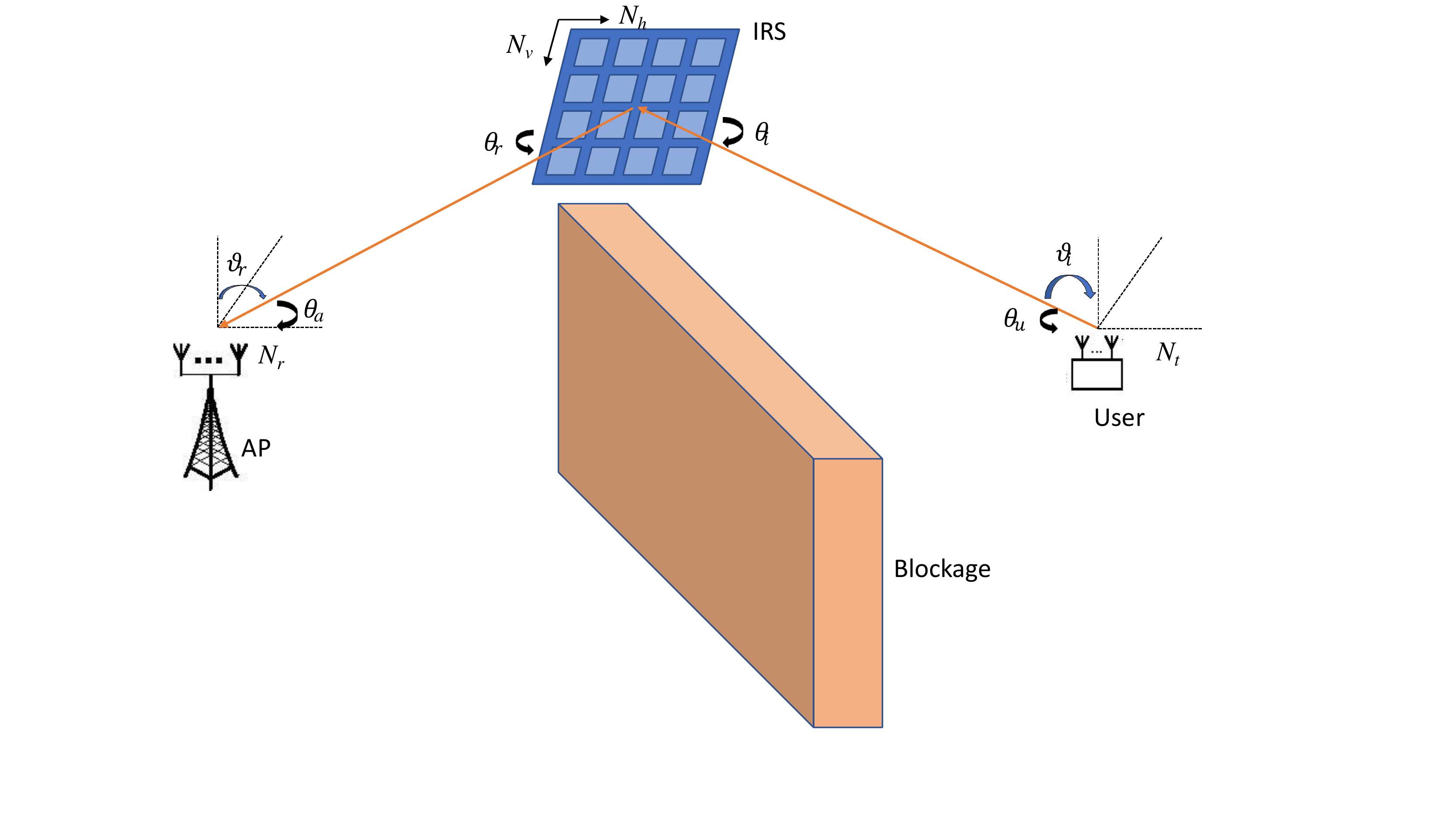}
\vspace{1.5em}
\caption{Uplink communication assisted by an RIS}
\label{fig:sys_model}
\end{centering}
\end{figure}

As shown in fig.~\ref{fig:sys_model} , we consider the uplink communication and beam training procedure assisted by an RIS. We consider that AP knows about fixed direction of signal reception from RIS and RIS also knows the direction of reflected beam to AP because AP is fixed. In this model, direction of incoming incident signal for RIS is not fixed because the user node is mobile and the direction of incident beam depends on location of user node. We consider a uniform linear array (ULA) of $N_t$ and $N_r$ antennas placed horizontally with a spacing of half wavelength at the user node and AP respectively.  We consider that a single RF chain is connected to ULA at both user node and AP and thus analog beamforming structure is considered. Analog beamforming is considered as one of the branches of the hybrid precoding structure hence the proposed codebook is applicable to the hybrid precoding structure. At the transmitter side each antenna is connected to a power amplifier (PA) and a phase shifter while at the receiver node each antenna is connected to a low noise amplifier (LNA) and a phase shifter. Generally PA and LNA have the same scaling factor when the antenna is activated. The RIS has $N_v$ rows and $N_h$ columns of reflecting elements placed at half wavelength i.e. $\frac{\lambda}{2}$. Hence, RIS is a planner rectangular array of  $N_I=N_h \times N_v$ reflecting elements.

LoS transmission makes an azimuth angle of $\theta_u$ with ULA at the user node. $\theta_i$ and $\vartheta_i$ are azimuth and elevation angle respectively for the incident beam at the RIS. The azimuth angle can vary from user to user depending on location of users. The RIS controller has to estimate this azimuth angle $\hat{\theta}_i$ for any user. Similarly the user is also mobile hence the user node has to estimate the azimuth angle $\hat{\theta}_u$ to initiate communication with AP through RIS. As the AP is fixed and its location is known to the RIS controller therefore the azimuth angle $\theta_r$ and the elevation angle $\pi-\vartheta_r$ of reflected beam are known to the RIS controller. The AP also knows the location of the RIS hence it tunes the phase shift of antenna elements for the azimuth angle $\theta_a$ and the elevation angle $\vartheta_r$ for LoS communication.

For a horizontally placed ULA of $N$ antenna elements with spacing $d$ and steering at an azimuth angle $\theta$ and an elevation angle $\vartheta$, the steering vector $u_h\left(N,\Omega\right)$ can be defined as 
\begin{align}
& \bm{u_h}\left(N,\Omega\right) = \frac{1}{\sqrt{N}}\left[e^{j\pi0\Omega},e^{j\pi1\Omega},e^{j\pi2\Omega},\ldots,e^{j\pi(N-1)\Omega}\right]^T, \\ 
&\mbox{where }  
\Omega = \frac{2d}{\lambda}\cos(\theta)\sin(\vartheta), \mbox{ }  \Omega \in [-1,1] \mbox{ for } d=\frac{\lambda}{2}. \nonumber  \label{eq:hsteering}
\end{align}
The horizontally placed ULA controls the steering of azimuth angle while the vertically placed ULA controls the elevation angle. Hence, for a vertically placed ULA of $N$ antenna elements with spacing $d$ and steering at an elevation angle $\vartheta$, the steering vector $u_v\left(N,\Phi\right)$ can be defined as 
\begin{align}
& \bm{u_v}\left(N,\Phi\right) = \frac{1}{\sqrt{N}}\left[e^{j\pi0\Phi},e^{j\pi1\Phi},e^{j\pi2\Phi},\ldots,e^{j\pi(N-1)\Phi}\right]^T,\\ 
&\mbox{where }  
\Phi = \frac{2d}{\lambda}\cos(\vartheta), \mbox{ }  \Phi \in [-1,1] \mbox{ for } d=\frac{\lambda}{2}. \nonumber  \label{eq:vsteering}
\end{align}

Steering vector at the RIS for the incident beam is given by 
\begin{align}
& \bm{u_i}\left(N_h,N_v,\Omega_i,\Phi_i\right) = \bm{u_h}\left(N_h,\Omega_i\right)\otimes \bm{u_v}\left(N_v,\Phi_i\right),\\ 
&\mbox{where }  
\Omega_i = \cos(\theta_i)\sin(\vartheta_i) \mbox{ and }\Phi_i = \cos(\vartheta_i).  \nonumber  \label{eq:isteering}
\end{align}
Here, $\otimes$ stands for the Kronecker product.
Similarly, the steering vector at the RIS for the reflected beam is given by 
\begin{align}
& \bm{u_r}\left(N_h,N_v,\Omega_r,\Phi_r\right) = \bm{u_h}\left(N_h,\Omega_r\right)\otimes \bm{u_v}\left(N_v,\Phi_r\right),\\ 
&\mbox{where }  
\Omega_r = \cos(\theta_r)\sin(\vartheta_r) \mbox{ and }\Phi_r = \cos(\vartheta_r). \nonumber  \label{eq:rsteering}
\end{align}

The channel matrix $\bm{H}$ between the RIS and the user node is given by
\begin{align}
& \bm{H} =\sqrt{N_IN_t} \bm{u_i}\left(N_h,N_v,\Omega_i,\Phi_i\right)\mu_{ui} \bm{u_h}\left(N_t,\Omega_u\right)^H, \\
&\mbox{where }  
\Omega_u = \cos(\theta_u)\sin(\vartheta_i). \nonumber  \label{eq:ichannel}
\end{align}
$\mu_{ui}$ is channel coefficient of superposition of LoS and NLoS components between the user node and the RIS.
The channel matrix $\bm{G}$ between the AP and the RIS is given by
\begin{align}
& \bm{G} =\sqrt{N_rN_I} \bm{u_h}\left(N_r,\Omega_a\right) \mu_{ia} \bm{u_r}\left(N_h,N_v,\Omega_r,\Phi_r\right)^H, \\
&\mbox{where }  
\Omega_a = \cos(\theta_a)\sin(\vartheta_r). \nonumber  \label{eq:rchannel}
\end{align}
$\mu_{ia}$ is the channel coefficient of superposition of LoS and NLoS components between the RIS and the AP. $\mu_{ui}$ and $\mu_{ia}$ are Rician fading channel gains.

$\bm{\Psi} \triangleq \mbox{diag}\left(\bm{\xi}\right)\in \mathbb{C}^{N_I{\times}N_I}$ is a passive beamforming matrix predicted by the RIS controller, where  $\bm{\xi} \triangleq [e^{j\pi\psi_1},e^{j\pi\psi_2}, \ldots, e^{j\pi\psi_{N_I}}]^T$. Further beamforming vector can be decoupled as $\bm{\xi}=\bm{\xi_h} \otimes \bm{\xi_v}$, where $\bm{\xi_h}$ is the  beamforming for the azimuth angle and $\bm{\xi_v}$ is the beamforming for the elevation angle to search the user node. The RIS controller has information about the direction of reflected beam and the elevation angle of the incident beam and hence, it can use this information in determining $\bm{\xi}$. $\bm{\xi_h}$ is the prediction of incident beam azimuth angle by the RIS controller following the procedure given in the section~ \ref{sec:algo}. Without loss of generality, we consider $\bm{w_u}$ and $\bm{w_a}$ as normalised antenna wave vectors (AWVs), i.e. $||\bm{w_u}||=||\bm{w_a}||=1$ at the user node and AP respectively. Note that the AP has information about the  location of RIS and hence $\bm{w_a}$ is fixed while user node has to search for RIS therefore $\bm{w_u}$ keeps on updating according to the procedure defined in the section~ \ref{sec:algo}. The beam width can be controlled by activating/deactivating antenna elements at the user node. Because the location of RIS is not known to the user node hence initially beam width should be large to cover the entire angular space. Let $N_{tact}$ represents the number of active radiating elements and $P_{per}$ is the power of each radiating element. AWVs can be obtained as follows:
\begin{align}
\bm{w_u} &=\bm{u_h}\left(N_{tact},{\Omega}'_u\right),   \\
\bm{w_a} & = \bm{u_h}\left(N_r,\Omega_a\right).\nonumber  \label{eq:awv}
\end{align}

Here ${\Omega}'_u$ is the predicted azimuth angle by the user node to search the RIS. This angle is updated using a codebook. User node transmits training signal $x$ and it reaches to the AP via RIS. Received signal at the AP is given by:

\begin{equation}
y =\sqrt{P_{per}N_{tact}}\bm{w_a}^H\bm{G}\bm{\Psi}\bm{H}\bm{w_u}x+\bm{w_a}^Hn, \label{eq:channelmodel}
\end{equation}
where $n$ is the additive white Gaussian noise (AWGN) with power $N_0$ i.e. $\mathbbm{E}(nn^H)=N_0\bm{I}$. Total transmission signal to noise ratio (SNR) is defined as $\gamma_{tot}=P_{tot}/N_0$, where $P_{tot}=P_{per}N_{tact}$. Received SNR at the AP is given by:

\begin{align}
\eta_{tot} &=\gamma_{tot}||\bm{w_a}^H\bm{G}\bm{\Psi}\bm{H}\bm{w_u}||^2 \nonumber \\ 
&=\gamma_{tot}{N_rN_t}||\mu_{ui}\mu_{ia} \sqrt{N_I}\bm{u_r}\left(N_h,N_v,\Omega_r,\Phi_r\right)^H \odot \sqrt{N_I}\bm{u_i}\left(N_h,N_v,\Omega_i,\Phi_i\right)^T\bm{\xi}\bm{u_h}\left(N_t,\Omega_u\right)^H\bm{w_u}||^2,\label{eq:SNR}\\
&=\gamma_{tot}{N_rN_t}||\mu_{ui}\mu_{ia} \bm{q}^H\bm{\xi}\bm{u_h}\left(N_t,\Omega_u\right)^H\bm{w_u}||^2. \nonumber 
\end{align}

Here, $\bm{q}^H= \sqrt{N_I}\bm{u_r}\left(N_h,N_v,\Omega_r,\Phi_r\right)^H \odot \sqrt{N_I}\bm{u_i}\left(N_h,N_v,\Omega_i,\Phi_i\right)^T$ and $\odot$ stands for the Hadamard product. Objective of this work is to propose an appropriate codebook for the AWV $\bm{w_u}$ , the passive beamforming vector $\bm{\xi}$ and a fast beam training algorithm to maximize $\eta_{tot}$. Further,

\begin{align}
\bm{q}^H &= \sqrt{N_I}\bm{u_r}\left(N_h,N_v,\Omega_r,\Phi_r\right)^H \odot \sqrt{N_I}\bm{u_i}\left(N_h,N_v,\Omega_i,\Phi_i\right)^T \nonumber \\
&= \sqrt{N_I}\left( \bm{u_h}\left(N_h,\Omega_r\right)\otimes \bm{u_v}\left(N_v,\Phi_r\right)\right)^H \odot \sqrt{N_I}\left(\bm{u_h}\left(N_h,\Omega_i\right)\otimes \bm{u_v}\left(N_v,\Phi_i\right)\right)^T \nonumber\\
&= \sqrt{N_h}\left( \bm{u_h}\left(N_h,\Omega_r\right)^H\odot \bm{u_h}\left(N_h,\Omega_i\right)^T\right) \otimes \sqrt{N_v}\left(\bm{u_v}\left(N_v,\Phi_r\right)^H\odot \bm{u_v}\left(N_v,\Phi_i\right)^T\right) \nonumber\\
&=  \sqrt{N_h}\bm{u_h}\left(N_h,\overline{\Omega}\right)^H \otimes \sqrt{N_v}\bm{u_v}\left(N_v,\overline{\Phi}\right)^H,
\label{eq:simplify}
\end{align}
where $\overline{\Omega}={\Omega}_r-{\Omega}_i$, $\overline{\Omega}\in[-2,2]$ and $\overline{\Phi}={\Phi}_r-{\Phi}_i$, $\overline{\Phi}\in[-2,2]$ are the phase shifts to be induced by the RIS. Note that $u_h\left(N,\Omega\right)$ and $u_v\left(N,\Phi\right)$ are periodic with period 2. Denote $\tilde{\Omega}=\overline{\Omega}\mbox{ (mod 2)}\in[-1,1]$ and $\tilde{\Phi} =\overline{\Phi}\mbox{ (mod 2)}\in[-1,1]$. 
Thus, $\bm{q}^H \bm{\xi}$ can be written as
\begin{align}
\bm{q}^H \bm{\xi} &=\left(\sqrt{N_h}\bm{u_h}(N_h,\tilde{\Omega})^H \otimes \sqrt{N_v}\bm{u_v}(N_v,\tilde{\Phi})^H \right)\left( \bm{\xi_h} \otimes \bm{\xi_v}\right) \nonumber \\
&=\left(\sqrt{N_h}\bm{u_h}(N_h,\tilde{\Omega})^H \bm{\xi_h}  \right)\otimes \left(\sqrt{N_v} \bm{u_v}(N_v,\tilde{\Phi})^H  \bm{\xi_v}\right).
\label{eq:simplify1}
\end{align}
For the optimal beam alignment at RIS, $\bm{\xi_h}=\sqrt{N_h}u_h(N_h,\tilde{\Omega})$ and $\bm{\xi_v}=\sqrt{N_v}u_v(N_v,\tilde{\Phi})$. Hence the RIS controller uses a codebook to predict $\tilde{\Omega}$ and get feedback from AP about the received SNR. Similarly the user node uses the codebook to maximize $\bm{u_h}\left(N_t,\Omega_u\right)^H\bm{w_u}$. Let $\tilde{\Omega}'$ and ${\Omega}'_u$ are the predictions made by the RIS controller and the user node respectively. Then the received SNR at the AP is 

\begin{eqnarray} \label{eq:SNR1}
\begin{split}
\eta_{tot} (\tilde{\Omega}',{\Omega}'_u)&=\gamma_{tot}{N_rN_t}\\
&||\mu_{ui}\mu_{ia} \bm{q}^H\left(\bm{u_h}(N_h,\tilde{\Omega}') \otimes \bm{u_v}(N_v,\tilde{\Phi})\right)\bm{u_h}\left(N_t,\Omega_u\right)^H \bm{u_h}\left(N_t,{\Omega}'_u\right)||^2. \nonumber 
\end{split}
\end{eqnarray}
Again written as
\begin{equation}
\eta_{tot} (\bm{\xi_h},\bm{w_u})=\gamma_{tot}{N_rN_t}
||\mu_{ui}\mu_{ia} \bm{q}^H\left( \bm{\xi_h} \otimes \bm{\xi_v} \right)\bm{u_h}\left(N_t,\Omega_u\right)^H \bm{w_u}||^2, \label{eq:SCAN}
\end{equation}
where $\bm{\xi_h}=\sqrt{N_h}\bm{u_h}(N_h,\tilde{\Omega}')$ and $\bm{w_u}=\bm{u_h}\left(N_t,\Omega'_u\right)$.
Thus, to obtain the proper beam alignment,
 it suffices to solve the
following optimization program:

\noindent
(O) ${\rm Maximize}_{\tilde{\Omega}',{\Omega}'_u} : \eta_{tot}(\tilde{\Omega}',{\Omega}'_u)$

\noindent
{\rm Subject to}: 

(1)  $ \tilde{\Omega}'\in [-1,1]$,

(2)  ${\Omega}'_u \in [-1,1]$.

Finding the solution of the above optimization problem by training in continuous range of directions is more accurate but it has more computational complexity therefore the codebook based switching beamforming training is a feasible solution for practical applications. Codebook based switching beamforming training actually discretize the range $[-1,1]$ in $N_h$ and $N_t$ different levels at the RIS and the user node respectively. Without loss of generality, we consider $ \tilde{\Omega}'\in \{\tilde{\Omega}^1,\ldots,\tilde{\Omega}^{N_h}\}$ and ${\Omega}'_u\in \{{\Omega}_u^1,\ldots,{\Omega}_u^{N_t}\}$.
\begin{definition}
	A beam training policy $\Delta$ is a search method for those possible pair sets $\{\tilde{\Omega}',{\Omega}'_u\}$  from all the possible $N_t \times N_h$ sets that maximize the received SNR $\eta_{tot}(\tilde{\Omega}',{\Omega}'_u)$.
\end{definition}
Let ${\tilde{\Omega}^{\Delta}},({\Omega}_u^{\Delta}, resp)$ be the prediction of the incident azimuth angle at the RIS (user node, resp) under policy $\Delta$. Then using \eqref{eq:SNR1}, the received SNR at the AP is given by $\eta_{tot}(\tilde{\Omega}^\Delta,{\Omega}_u^\Delta)$.
\begin{definition}
	A beam training policy $\Delta$ is said to be feasible if
	it maximizes the received SNR at the AP in a given time constraint.
\end{definition}

Thus, a feasible beam training policy respects the time constraint in setting a virtual LoS communication link between the user node and the AP for practical applications. It quickly scans all the possible directions in the sets $ \{\tilde{\Omega}^1,\ldots,\tilde{\Omega}^{N_h}\}$, $\{{\Omega}_u^1,\ldots,{\Omega}_u^{N_t}\}$ and selects the best pair $\{\tilde{\Omega}',{\Omega}'_u\}$. 

To the best of our knowledge, literature does not discuss beam training for the random access from the user node for the RIS assisted communication. Next we discuss the codebook based approach for beam training at the RIS and the user node.

\section{Proposed beam training algorithm}
\label{sec:algo}
This section is divided into two parts. In the first part, we propose beam training at the RIS while user node transmits using an omni directional antenna. Once RIS controller get passive beamforming by tuning the RIS, beam training is obtained at the user node by tuning its antenna array. An optimal beam training policy $\Delta^*$ must solve the following optimization program with minimum number of training sweep symbols for practical applications:

\noindent
(O) ${\rm Maximize}_{\tilde{\Omega}',{\Omega}'_u} : \eta_{tot}(\tilde{\Omega}',{\Omega}'_u)$

\noindent
{\rm Subject to}: 

(1)  $ \tilde{\Omega}'\in \{\tilde{\Omega}^1,\ldots,\tilde{\Omega}^{N_h}\}$,

(2)  ${\Omega}'_u \in \{{\Omega}_u^1,\ldots,{\Omega}_u^{N_t}\}$.

Here we use structural properties of the problem to obtain the beam training policy $\Delta$. It must be noticed that the beam training at RIS and the user node can be decoupled and hence problem can be divided into two sub-problems. Proposed algorithm used alternating maximization approach to maximize SNR at the AP. First we fix the active beamforming at the user node and obtain passive beamforming at the RIS for the directions of transmission that maximizes the SNR at the AP. Later we fix the passive beamforming at the RIS and obtain active beamforming at the user node for the direction which maximize SNR at the AP. Hence variables $\{\tilde{\Omega}',{\Omega}'_u\}$ are divided in two parts. Then we solve the following optimization programs

\noindent
(OI) ${\rm Maximize}_{\tilde{\Omega}'} : \eta_{tot}(\tilde{\Omega}',{\Omega}'_u)$

\noindent
{\rm Subject to}: 

$\bullet$   $ \tilde{\Omega}'\in \{\tilde{\Omega}^1,\ldots,\tilde{\Omega}^{N_h}\}$, and

\noindent
(OU) ${\rm Maximize}_{{\Omega}'_u} : \eta_{tot}(\tilde{\Omega}',{\Omega}'_u)$

\noindent
{\rm Subject to}: 

$\bullet$   ${\Omega}'_u \in \{{\Omega}_u^1,\ldots,{\Omega}_u^{N_t}\}$.

To reduce the number of sweeps we propose a codebook for $\tilde{\Omega}'$ and ${\Omega}'_u$.
\subsection{Codebook for beam training at RIS}
\label{Codebook_IRS}
\begin{figure}[h]
\vspace{1em}
\begin{centering}
\includegraphics[scale=0.5]{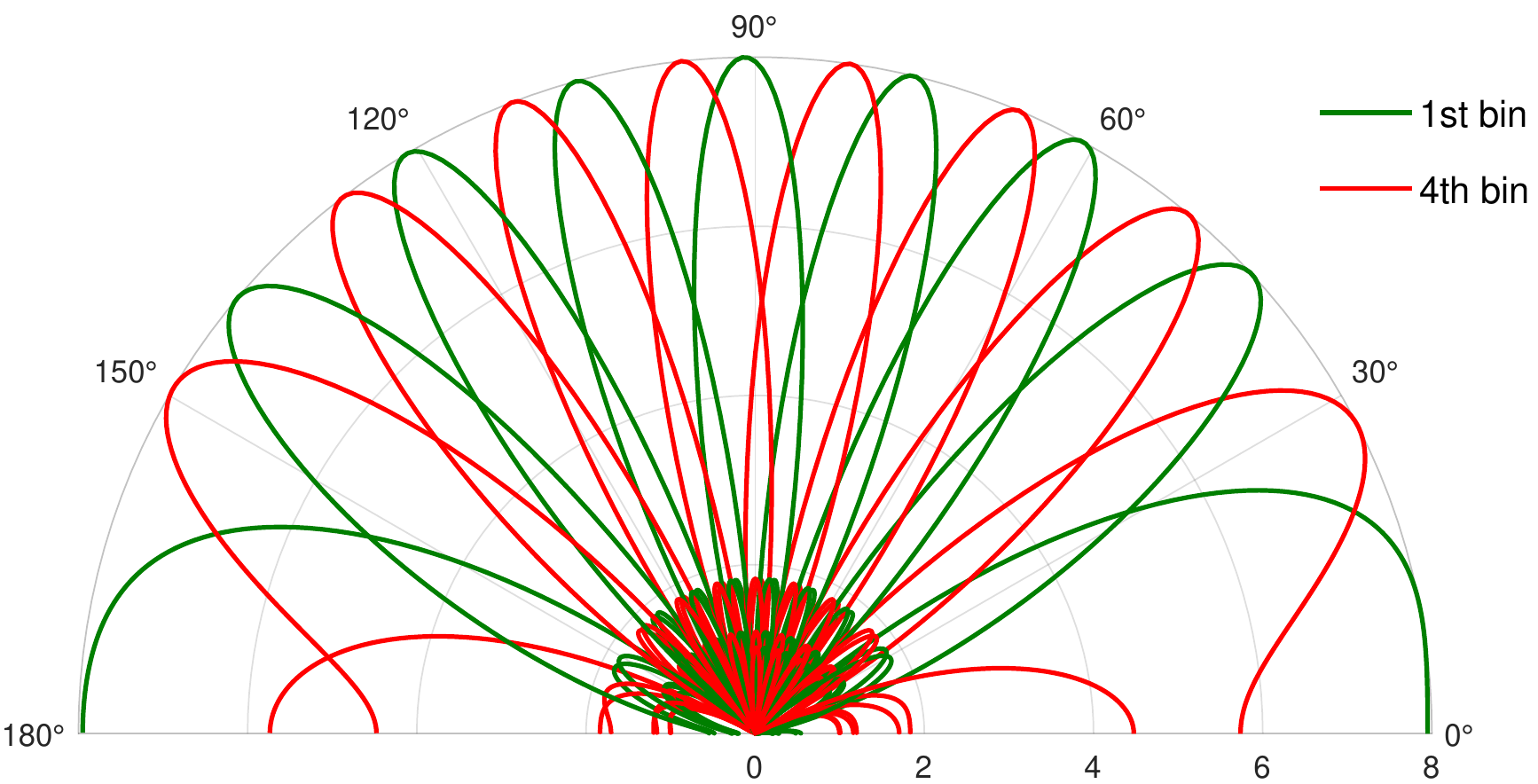}
\vspace{1.5em}
\caption{Beams of $1st$ bin and $4th$ bin of training codebook with $M=8$ and $L=8$}
\label{fig:IRS_code}
\end{centering}
\end{figure}

Throughout this section, we have considered that the user node is transmitting at fixed direction ${\Omega}'_u$. We do beam training at the RIS by following a codebook designed for RIS controller. As we have considered semi passive RIS hence the AP provides information of received SNR to the RIS controller as a feedback to follow a proper sequence in codebook. To minimize the number of sweeps, we follow a multi beam training codebook at the RIS.  The expected incident angle space $\{{\Omega}_i^1,\ldots,{\Omega}_i^{N_h}\}$ has a unique mapping in phase shift space of RIS given by $\{\tilde{\Omega}^1,\ldots,\tilde{\Omega}^{N_h}\}$. First we divide entire phase shift direction space in $N_h$ equal size sectors with central direction given by $\tilde{\Omega}^j \mbox{ where } j\in \mathcal{J} \triangleq \{1,\ldots,N_h\}$.  Hence $\tilde{\Omega}^j$ is element $j$ in $\{\tilde{\Omega}^1,\ldots,\tilde{\Omega}^{N_h}\}$ and $\tilde{\Omega}^j=\left(-1+\frac{2j-1}{N_h}\right)$. For any beam direction set $\mathcal{A} \subseteq \mathcal{J}$, we define an \textit{intra-set distance} as $d_s(\mathcal{A})=\mbox{min}_{i,j \in\mathcal{A};i\neq j}|i-j|$. The RIS controller sweeps $M$ different directions in a single sweep. Hence $N_h$ elements of a row on RIS are subdivided into $M$ different sub-arrays where each sub-array consists of $L \triangleq \frac{N_h}{M}$ elements. Hence $L$ bins cover complete space $\{\tilde{\Omega}^1,\ldots,\tilde{\Omega}^{N_h}\}$. Therefore, the direction space $\{\tilde{\Omega}^1,\ldots,\tilde{\Omega}^{N_h}\}$ is divided in to $L$ bins belonging to the set $\mathcal{B}=\{B(1),\ldots,B(L)\}$, where bin $B(l)$ consists of directions $\{\beta_1^l,\ldots,\beta_M^l\}$ and $\beta_m^l$  represents  direction $m$ of bin $l$ and $\beta_m^l=\tilde{\Omega}^{l+(m-1)L}\,\, \forall l \in \{1,\ldots,L\} \mbox{ and } m \in \{1,\ldots,M\}$. Such allocation of directions in a bin actually maximize the intra bin distance to minimize the interference between adjacent beams within a bin as shown in Table~\ref{tab:codebook}.  Steering vector for bin $B(l)$ is $\bm{\tilde{W}_{1,l}}=[\bm{\tilde{w}_{1,1}}^T(\beta_1^l),\ldots,\bm{\tilde{w}_{1,M}}^T(\beta_M^l)]^T$, where $\bm{\tilde{w}_{1,m}}(\alpha)\triangleq \bm{u_h}(N_h,\alpha)_{(m-1)L+1:mL} =e^{j(m-1)L\alpha}\bm{u_h}(L,\alpha)$. Note that multi beam sweeping using $L$ radiating elements have broad beams while smaller gains compared to single beam scan using $N_h$ radiating elements. First round of beam sweeping consider all the bins hence $\bm{\xi_h}=\bm{\tilde{W}_{1,l}}\,\,\forall l \in \{1,\ldots,L\}$ in equation \eqref{eq:SCAN}. Hence for first round of sweep $\tilde{S}(1,l)=B(l)\,\,\forall l \in \{1,\ldots,L\}$. Note that for first round of sweep, intra bin distance is $d_s(B(l))=L \,\,\forall l\in \{1,\ldots,L\}$ (see Table~\ref{tab:codebook}). In case of multi beam training chances of false bin detection increases with the increase in $M/L$ as bins having directions nearest to desired direction raises SNR due to widening of beams with the high value of $M$ (see figure~\ref{fig:IRS_code}). Therefore the error in first step can be corrected by selecting $T$ number of bins after the first round of sweep. The value of $T$ depends on $M/L$ ratio. Generally $T$ is higher for the high value of  $M/L$. These $T$ bins are those bins which yield high SNR in the first round of beam sweeping. These $T$ bins are used in the fine search after the last stage of coarse search to minimize error of detection from the first round of beam sweeping (see Table~\ref{tab:codebook4}). It must be noted that the different bins have nearest directions at same index which leads to beam overlapping with the increase in $M$ and hence increase the probability false detection of bin after the first round of beam sweeping (see Table~\ref{tab:codebook}). Let $\eta(r,B(l))$ be the received SNR at the AP from bin $l$ in round $r$, initially set $\mathcal{G}(0)= \varnothing,\,\mathcal{P}=\mathcal{B},\, \texttt{prev}=0,\,\texttt{next}=1$ and after the first round of sweep, get $\mathcal{G}(T)$ by repeating $\mathcal{G}(\texttt{next})=\{\mathcal{G}(\texttt{prev}), \mathcal{T}\} ,\mbox{ where }\mathcal{T}=\mbox{arg max}_{B(l)\in \mathcal{P}}\eta(1,B(l))$ and update 
$\mathcal{P}=\mathcal{P}\backslash \mathcal{T}$, $\,\texttt{prev}=\texttt{next},\,\, \texttt{next}=\texttt{next}+\mbox{size}(\mathcal{T})$ till $\texttt{next}\le L$, here $\mbox{size}(\mathcal{T})$ represents the number of bin in set $(\mathcal{T})$. Finally, $\mathcal{G}(L)=\{B'(1),\ldots,B'(L)\}$, where $\eta(1,B'(1))\le\eta(1,B'(2))\le \ldots \le\eta(1,B'(L))$ (see Table~\ref{tab:codebook1}). Therefore $\mathcal{G}(T)=[\mathcal{G}(L)]_{1:T}$ is obtained by sorting bins of $\mathcal{B}$ in the descending order of received SNR and selecting the first $T$ bins from the ordered set.
\begin{table}[h]
\caption{Codebook for muti-beam training with $N_h=64$, $M=8$ and $L=8$.}\label{tab:codebook}
\centering
    \begin{tabular}{l|c|c|c|c|c|c|c|c|}
    \cline{2-9}
    \hfil $B(1)$& \hfil {$1$} & \hfil {$9$} & \hfil {$17$} & \hfil {$25$} & \hfil {$33$} & \hfil {$41$} & \hfil {$49$} & \hfil {$57$}\\ \cline{2-9}
    \hfil $B(2)$& \hfil {$2$} & \hfil {$10$} & \hfil {$18$} & \hfil {$26$} & \hfil {$34$} & \hfil {$42$} & \hfil {$50$} & \hfil {$58$}\\ \cline{2-9}
    \hfil $B(3)$& \hfil {$3$} & \hfil {$11$} & \hfil {$19$} & \hfil {$27$} & \hfil {$35$} & \hfil {$43$} & \hfil {$51$} & \hfil {$59$}\\ \cline{2-9}    

    \hfil $B(4)$& \hfil {$4$} & \hfil {$12$} & \hfil {$20$} & \hfil {$28$} & \hfil {$36$} & \hfil {$44$} & \hfil {$52$} & \hfil {$60$} \\\cline{2-9}
    \hfil $B(5)$& \hfil {$5$} & \hfil {$13$} & \hfil {$21$} & \hfil {$29$} & \hfil {$37$} & \hfil {$45$} & \hfil {$53$} & \hfil {$61$} \\ \cline{2-9}
    \hfil $B(6)$& \hfil {$6$} & \hfil {$14$} & \hfil {$22$} & \hfil {$30$} & \hfil {$38$} & \hfil {$46$} & \hfil {$54$} & \hfil {$62$} \\ 
     \cline{2-9}
    \hfil $B(7)$& \hfil {$7$} & \hfil {$15$} & \hfil {$23$} & \hfil {$31$} & \hfil {$39$} & \hfil {$47$} & \hfil {$55$} & \hfil {$63$}\\ \cline{2-9}
    \hfil $B(8)$& \hfil {$8$} & \hfil {$16$} & \hfil {$24$} & \hfil {$32$} & \hfil {$40$} & \hfil {$48$} & \hfil {$56$} & \hfil {$64$}\\ \cline{2-9}    
    \end{tabular}
\end{table}

\begin{table}[h]
\caption{Bins of codebook arranged in descending order of SNR.}\label{tab:codebook1}
\centering
    \begin{tabular}{l|c|c|c|c|c|c|c|c|}
    \cline{2-9}
    \hfil $B'(1)$& \hfil {$6$} & \hfil {$14$} & \hfil {$22$} & \hfil {$30$} & \hfil {$38$} & \hfil {$46$} & \hfil {$54$} & \hfil {$62$} \\ 
     \cline{2-9}
    \hfil $B'(2)$& \hfil {$5$} & \hfil {$13$} & \hfil {$21$} & \hfil {$29$} & \hfil {$37$} & \hfil {$45$} & \hfil {$53$} & \hfil {$61$} \\ \cline{2-9}
    \hfil $B'(3)$& \hfil {$4$} & \hfil {$12$} & \hfil {$20$} & \hfil {$28$} & \hfil {$36$} & \hfil {$44$} & \hfil {$52$} & \hfil {$60$} \\ \cline{2-9}
    \hfil $B'(4)$& \hfil {$2$} & \hfil {$10$} & \hfil {$18$} & \hfil {$26$} & \hfil {$34$} & \hfil {$42$} & \hfil {$50$} & \hfil {$58$}\\ \cline{2-9}
    \hfil $B'(5)$& \hfil {$7$} & \hfil {$15$} & \hfil {$23$} & \hfil {$31$} & \hfil {$39$} & \hfil {$47$} & \hfil {$55$} & \hfil {$63$}\\ \cline{2-9}
    \hfil $B'(6)$& \hfil {$3$} & \hfil {$11$} & \hfil {$19$} & \hfil {$27$} & \hfil {$35$} & \hfil {$43$} & \hfil {$51$} & \hfil {$59$}\\ \cline{2-9}
    \hfil $B'(7)$& \hfil {$1$} & \hfil {$9$} & \hfil {$17$} & \hfil {$25$} & \hfil {$33$} & \hfil {$41$} & \hfil {$49$} & \hfil {$57$}\\ \cline{2-9}
    \hfil $B'(8)$& \hfil {$8$} & \hfil {$16$} & \hfil {$24$} & \hfil {$32$} & \hfil {$40$} & \hfil {$48$} & \hfil {$56$} & \hfil {$64$}\\ \cline{2-9}    
    \end{tabular}
\end{table}

\begin{table}[h]
\caption{Bin with highest SNR for successive rounds of sweep.}\label{tab:codebook2}
\centering
    \begin{tabular}{l|c|c|c|c|c|c|c|c|}
    \cline{2-9}
    \hfil $B'(1)$& \hfil {$6$} & \hfil {$14$} & \hfil {$22$} & \hfil {$30$} & \hfil {$38$} & \hfil {$46$} & \hfil {$54$} & \hfil {$62$} \\ 
     \cline{2-9}    
    \end{tabular}
\end{table}

\begin{table}[h]
\centering
    \begin{tabular}{l|c|c|c|c|}
    \cline{2-5}
    \hfil $\tilde{S}(2,1)$& \hfil {$6$} & \hfil {$14$} & \hfil {$22$} & \hfil {$30$}  \\ 
     \cline{2-5}    
    \end{tabular}
\end{table}

\begin{table}[h]
    \caption{Expected directions after $log_2(M)$ rounds of sweep from bin corresponding to highest SNR.}\label{tab:codebook3}
\centering
    \begin{tabular}{l|c|c|}
    \cline{2-3}
    \hfil $\tilde{S}(log_2(M),1)$& \hfil {$38$} & \hfil {$46$}  \\ 
     \cline{2-3}    
    \end{tabular}
\end{table}

\begin{table}[h]
\caption{Expected directions after $log_2(M)+1$ for fine search.}\label{tab:codebook4}
\centering
    \begin{tabular}{l|c|c|c|c|}
    \cline{2-3} \cline{5-5}
    \hfil $[B'(1)]_{m:m+1}$& \hfil {$38$} & \hfil {$46$} &\hfil $D(1)$&  \hfil {$46$}  \\ \cline{2-3} \cline{5-5}
    \hfil $[B'(2)]_{m:m+1}$& \hfil {$37$} & \hfil {$45$}  &\hfil $D(2)$&  \hfil {$45$} \\ \cline{2-3} \cline{5-5}
    \hfil $[B'(3)]_{m:m+1}$& \hfil {$36$} & \hfil {$44$}  &\hfil $D(3)$&  \hfil {$44$} \\ \cline{2-3} \cline{5-5}
    \hfil $[B'(4)]_{m:m+1}$& \hfil {$34$} & \hfil {$42$}  &\hfil $D(4)$&  \hfil {$42$} \\ \cline{2-3} \cline{5-5}
    \hfil $[B'(5)]_{m:m+1}$& \hfil {$39$} & \hfil {$47$}  &\hfil $D(5)$&  \hfil {$47$} \\ \cline{2-3} \cline{5-5}
    \hfil $[B'(6)]_{m:m+1}$& \hfil {$35$} & \hfil {$43$}  &\hfil $D(6)$&  \hfil {$43$} \\ \cline{2-3} \cline{5-5}
    \hfil $[B'(7)]_{m:m+1}$& \hfil {$33$} & \hfil {$41$}  &\hfil $D(7)$&  \hfil {$41$} \\ \cline{2-3} \cline{5-5}
    \hfil $[B'(8)]_{m:m+1}$& \hfil {$40$} & \hfil {$48$}  &\hfil $D(8)$&  \hfil {$48$} \\ \cline{2-3} \cline{5-5}   
    \end{tabular}
\end{table}

\textit{(1) Coarse search: }
Coarse search involves beam sweeping with multiple beams. In the next round of a coarse search, we reduce the number of directions for beam sweeping by a factor of $1/2$, hence number of antennas used for sweeping a direction get doubled and it makes the beam narrower compared to the previous round. Therefore the probability of false detection reduces in the next round. For coarse search $r \in\{1,\ldots,log_2(M)\}$ and in the last round of coarse search i.e. $r=log_2(M)$, number of directions covered in a sweep are two. Let $\tilde{X}(r)$ represents the set of directions providing highest SNR at the AP after round $r$, then $\tilde{X}(1)=B'(1)=B(k),\eta_{max}=\eta(1,B(k)),\mbox{ where }B(k)=\mbox{arg max}_{B(l)\in \mathcal{B}}\eta(1,B(l))$ (see Table~\ref{tab:codebook2}). Hence for round $r$, $\tilde{S}(r,1)=[\tilde{X}(r-1)]_{1:x(r)}\,\,\forall r \in \{2,\ldots,log_2(M)\}\mbox{ where } x(r)=\frac{M}{2^{r-1}}$ (see Table~\ref{tab:codebook3}). Steering vector for sweep $\tilde{S}(r,1)$ in round $r$ is $\bm{\tilde{W}_{r,1}}=[\bm{\tilde{w}_{r,1}}^T(\alpha_1^1),\ldots,\bm{\tilde{w}_{r,x(r)}}^T(\alpha_{x(r)}^1)]^T$, where $\bm{\tilde{w}_{r,m}}(\alpha)\triangleq \bm{u_h}(N_h,\alpha)_{(m-1)(2^{r-1})L+1:m(2^{r-1})L} =e^{j(m-1)(2^{r-1})L\alpha}\bm{u_h}((2^{r-1})L,\alpha)$. Obtain $\eta_{th}=\eta_{max}/2$. Set consisting best beam direction in round  $r\in\{2,\ldots,log_2(M)\}$ after sweeping bin $\tilde{S}(r,1) $ is given by

\begin{equation}
  \tilde{X}(r) = \begin{cases}
  \tilde{X}(r-1) \cap \tilde{S}(r,1),  & \mbox{if } \,\, \eta(r,[\tilde{X}(r-1)]_{1:x(r)})>\eta_{th} \\
  \tilde{X}(r-1) \backslash \tilde{S}(r,1),  & \mbox{otherwise}   .
  \end{cases} \label{eq:update}\\
\end{equation}

After round $log_2(M)$, we get $\tilde{X}(log_2(M))=[\tilde{X}(log_2(M))(1),\tilde{X}(log_2(M))(2)]$.

\textit{(2) Fine search: }
Whole idea behind fine search is to correct the possibility of wrong bin selection in the first round due to high beam width with large $M$. Fine search uses single beam to identify the best direction. Fine search is done on the set $\mathcal{G}(T)$ that has been obtained after round $1$. First find $ m=\mbox{arg}_{k\in \{1,\ldots,M\}}\{B'(1)(k)=\tilde{X}(log_2(M))(1)\}$ then obtain $\{B'(l)(m),B'(l)(m+1)\} \,\,\forall l\in \{1,\ldots,T\}$. For round  $log_2(M)+1$ consider $\{B'(1)(m),B'(1)(m+1)\}$, let $\tilde{S}(log_2(M)+1,1)=B'(1)(m)$, if  $\eta(r,B'(1)(m))>\eta_{th}$ then set $k=m$ otherwise $k=m+1$. Set $p=B'(1)(k)$ and as shown in Table~\ref{tab:codebook4}, find $D(l)\,\,\forall l\in \{1,\ldots,T\}$ for the set of expected directions $D$ as 

\begin{equation}
  D(l) = \begin{cases}
  B'(l)(m) ,  & \mbox{if } \,\, |B'(l)(m)-p|<|B'(l)(m+1)-p| \\
  B'(l)(m+1) ,  & \mbox{if } \,\, |B'(l)(m)-p|>|B'(l)(m+1)-p| \\
B'(l)(k) ,  & \mbox{if } \,\, |B'(l)(m)-p|=|B'(l)(m+1)-p|.
  \end{cases}\label{eq:update1} \\
\end{equation}

For $r=log_2(M)+2$, $\tilde{S}(log_2(M)+2,l)$ uses a single direction beam sweeping vector $\bm{\tilde{W}_{r,l}}=\bm{u_h}(N_h,D(l))$. Desired direction of training RIS is $D(k)$, where $k=\mbox{arg max}_{l\in \{1,\ldots,T\}}\{\eta(log_2(M)+2,D(l))\}$. Therefore using codebook $\tilde{\Omega}'=\tilde{\Omega}^{D(k)}$ while optimal direction $\tilde{\Omega}_{opt}=\tilde{\Omega}^x$ where $x=\mbox{arg min}_{l\in \{1,\ldots,N_h\}}|\tilde{\Omega}^l-\tilde{\Omega}|$.

\subsection{Codebook for beam training at user node}
Once the RIS knows the location of the user node, passive beamforming training at the RIS is complete. Now the AP can send a signal to the user node using an RIS. In this downlink communication the user node performs beam training to maximize the received SNR. The user node can also make use of existing beam training codebook designed in last section for the RIS controller. With little abuse of notations, we consider same notation $w_u$ for receive beamforming at user node which is actually transmit beamforming vector in previous section. Similarly $w_a$ is transmit beamforming at the AP. For the downlink communication, the received SNR at the user node is given by 
\begin{align}
\eta_{tot} &=\gamma_{tot}||\bm{w_u}^H\bm{H}^H\bm{\Psi}^H\bm{G}^H\bm{w_a}||^2 \nonumber \\ 
&=\gamma_{tot}{N_rN_t}||\overline{\mu_{ui}\mu_{ia}} \bm{w_u}^H\bm{u_h}\left(N_t,\Omega_u\right)\bm{\xi}^H\bm{q}||^2. \nonumber 
\end{align}
Therefore corresponding equation \eqref{eq:SCAN} for the received SNR at the user node is 
\begin{equation}
\eta_{tot} (\bm{\xi_h},\bm{w_u})=\gamma_{tot}{N_rN_t}
||\overline{\mu_{ui}\mu_{ia}} \bm{w_u}^H\bm{u_h}\left(N_t,\Omega_u\right)\left( \bm{\xi_h} \otimes \bm{\xi_v} \right)^H\bm{q}||^2. \label{eq:SCAN1}
\end{equation}

It must be noted that maximizing the received SNR in equation~\eqref{eq:SCAN1} at the user node is equivalent to maximizing the SNR in equation~\eqref{eq:SCAN} for the transmit beamforming vector $\bm{w_u}$. Hence we consider the original objective function given in equation \eqref{eq:SCAN}. Therefore the codebook designed for the RIS (see subsection~\ref{Codebook_IRS}) also applied for beam training at the user node.

\begin{algorithm}
		\caption{Beam direction search algorithm}
		\label{fig:algorithm1}		
\hspace*{\algorithmicindent} \textbf{Input} Channel gains $\mu_{ui}, \mu_{ia}$, Parameters $\gamma_{tot},{N_r,N_t,N_v},M,L,N_h=M \times L, T,\bm{q}^H,\bm{\xi_v},\bm{u_h}\left(N_t,\Omega_u\right)$,
		expected phase shifts $\tilde{\Omega}^j=\left(-1+\frac{2j-1}{N_h}\right)\,\, \mbox{ where } j \in \{1,\ldots,N_h\}$ and $\mathcal{B}=\{B(1),\ldots,B(L)\}$ \\
\hspace*{\algorithmicindent} \textbf{Output} $\bm{\xi_h},\bm{w_u}$
\begin{algorithmic}[1]
\Procedure{Direction Search}{}				
\State 	Initialize: $r=1$, $\bm{w_u} = [\bm{u_h}\left(1,1\right)^T,\bm{0_{{(N_t-1)} \times 1}}^T]^T $
\While{$r \le log_2(M)+2$}\label{ln:first_while}
\If{$r=1$}
\State	Sweep $\tilde{S}(r,l)=B(l)\,\,\forall l \in \{1,\ldots,L\}$ and get $\eta(r,B(l))$.
\State Obtain $\mathcal{G}(T)=\{B'(1),\ldots,B'(T)\}$ where $\eta(1,B'(1))\le\eta(1,B'(2))\le \ldots \le\eta(1,B'(L))$
\State Initialize: $\hat{X}(1)=B'(1)$ and $\eta_{th}=\eta(1,B'(1))/2$
\EndIf
\If{$1<r \le log_2(M)$}
\State Obtain $\tilde{S}(r,1)=[\hat{X}(r-1)]_{1:x(r)}$ where $x(r)=\frac{M}{2^{r-1}}$.
\State Update $\hat{X}(r)$ as in~\eqref{eq:update}
\EndIf
\If{$r= log_2(M)+1$} 
\State Get $D(l)\,\,\forall l\in \{1,\ldots,T\}$ as in~\eqref{eq:update1}
\EndIf
\If{$r= log_2(M)+2$}
\State Sweep $\tilde{S}(r,l)=D(l))\,\,\forall l\in \{1,\ldots,T\}$
\State $\bm{\xi_h}=\bm{u_h}(N_h,D(k))$, where $k=\mbox{arg}_{l\in \{1,\ldots,T\}}\{\eta(r,D(l))\}$.
\EndIf
\EndWhile

\State 	Initialize: $r=1$, $M \mbox{ for user node }$ and $L=\frac{N_t}{M}$, expected directions $\hat{\Omega}^i_u=\left(-1+\frac{2i-1}{N_t}\right)\,\, \mbox{ where } i \in \{1,\ldots,N_t\}$ , $\mathcal{B}=\{B(1),\ldots,B(L)\}$, T \\
\While{$r \le log_2(M)+2$}\label{ln:second_while}
\If{$r=1$}
\State	Sweep $\tilde{S}(r,l)=B(l)\,\,\forall l \in \{1,\ldots,L\}$ and get $\eta(r,B(l))$.
\State Obtain $\mathcal{G}(T)=\{B'(1),\ldots,B'(T)\}$ where $\eta(1,B'(1))\le\eta(1,B'(2))\le \ldots \le\eta(1,B'(L))$
\State Initialize: $\hat{X}(1)=B'(1)$ and $\eta_{th}=\eta(1,B'(1))/2$
\EndIf
\If{$1<r \le log_2(M)$}
\State Obtain $\tilde{S}(r,1)=[\hat{X}(r-1)]_{1:x(r)}$ where $x(r)=\frac{M}{2^{r-1}}$.
\State Update $\hat{X}(r)$ as in~\eqref{eq:update}
\EndIf
\If{$r= log_2(M)+1$} 
\State Get $D(l)\,\,\forall l\in \{1,\ldots,T\}$ as in~\eqref{eq:update1}
\EndIf
\If{$r= log_2(M)+2$}
\State Sweep $\tilde{S}(r,l)=D(l))\,\,\forall l\in \{1,\ldots,T\}$
\State $\bm{w_u}=\bm{u_h}(N_t,D(k))$, where $k=\mbox{arg}_{l\in \{1,\ldots,T\}}\{\eta(r,D(l))\}$.
\EndIf
\EndWhile

\EndProcedure
\end{algorithmic}		
\end{algorithm}

\subsubsection{Complexity Analysis}
In this section, we do analysis over the complexity of the proposed beamforming training method. Based on search method discussed in the algorithm, the number of training symbols are given by 
\begin{equation}
  N_{TS}=L+log_2(M)+T+1.
\end{equation}
Here $L+log_2(M)+T+1$ are training symbols at the RIS. High value of $M$ reduces the number of training symbols but at the same time increases the probability of false detection hence to minimize that high value of $T \le L$ is considered.

\section{Numerical Evaluations}
\label{sec:ne}

This section discusses the simulation results to show the performance of the proposed algorithm. We consider a mmWave system with a carrier frequency of 30 GHz. Distance between RIS and user node is $d_{ui}=2$ meter and distance between AP and RIS is $d_{ia}=10$ meter. $N_h$ and $N_v$ at RIS are taken as $160$ and spacing between RIS elements is $\lambda/2$. Reference pathloss $\zeta_0$ is $-62 \text{ dB}$. Path loss exponent for user to RIS link is $\delta_{ui}=2.3$ and path loss exponent for RIS to BS link is taken as $\delta_{ia}=2$. Rician fading for AP-RIS $\kappa_{ia}$ is $5 \text{ dB}$ and Rician fading for User-RIS $\kappa_{ui}$ is $10 \text{ dB}$. Noise Power $N_0$ is $-109 \text{ dBm}$. We present the simulation results to validate the improvements in success probability due to proposed scheme (\emph{PS}) with correction in the later stage. First, we compare the results with coarse search (\emph{CS}) codebook for RIS without refinement \cite{Zhang2020}. Later in this chapter, we show the impact of a few key factors relevant to the beam alignment success probability, including the size of RIS, number of directions in a sweep, the number of bins chosen after round one and the signal to noise ratio. We define average SNR of RIS-assisted mmWave communication system as 
$SNR=\frac{P_{tot}(\zeta_0d_{ia}^{\delta_{ia}})(\zeta_0d_{ui}^{\delta_{ui}}){N_v^2}{N_h^2}{N_r}}{N_0}$, where $P_{tot}$ is the power of the transmitting node.  After getting $\xi_h \mbox{ and } w_u$ from the codebook, Rate $R$ can be obtained as $R=log_2(1+\eta_{tot} (\xi_h,w_u))$. If ${iter}$ represents total number of iterations, accuracy of beam identification is given by probability of success rate $P_{suc}=\frac{\sum_{k=1}^{iter}\mathbbm{1}(\tilde{\Omega}'=\tilde{\Omega}_{opt})}{iter}$, where $\mathbbm{1}(\bullet)$ stands for indicator function. For simulation results, we have averaged over 7500 different Rician fading channel realisations. 

\begin{figure}[h]
\vspace{1em}
\begin{centering}
\includegraphics[scale=0.5]{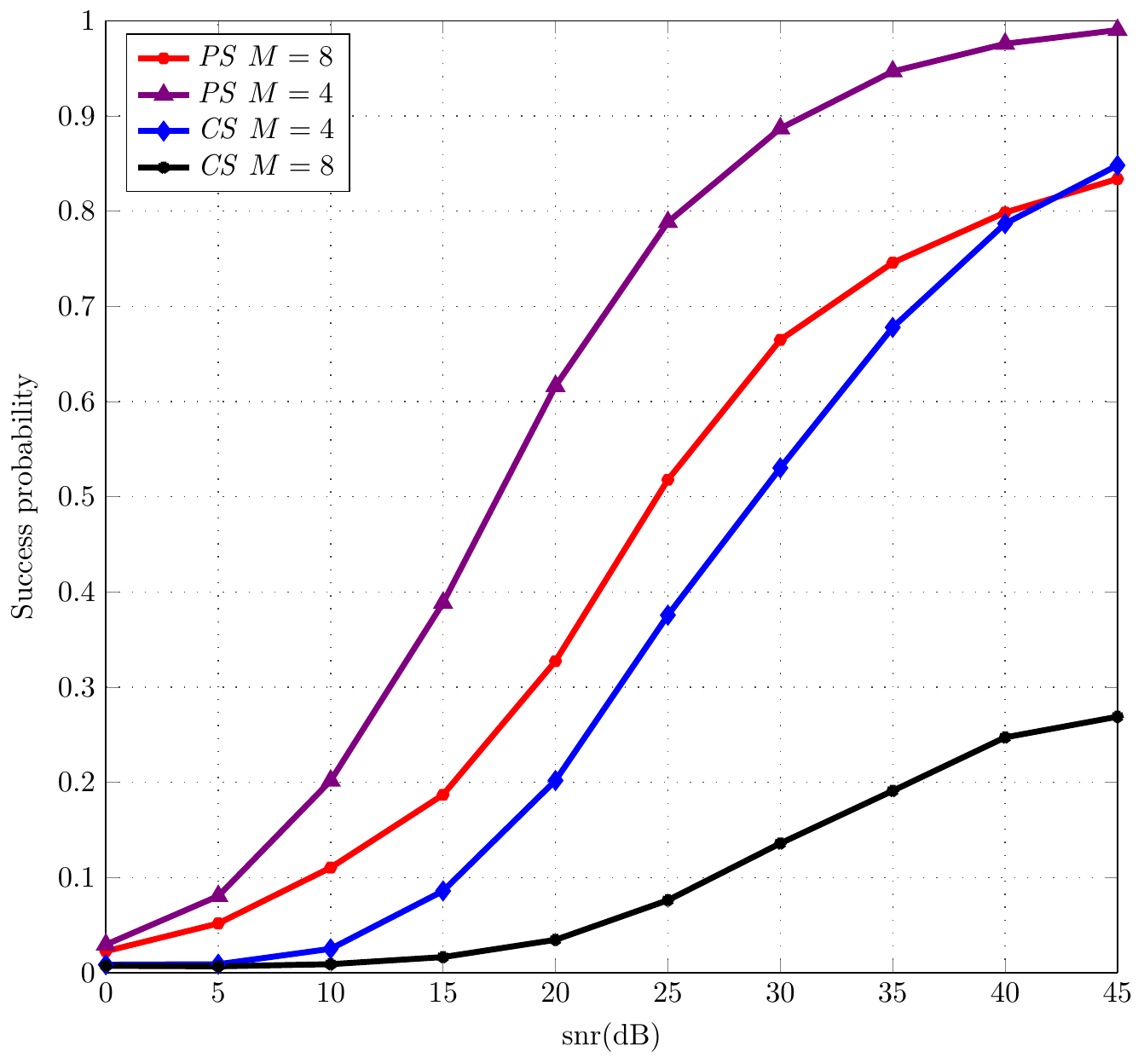}
\vspace{1.5em}
\caption{Success rate versus SNR for different number of sub-arrays at RIS}
\label{fig:sec_prob}
\end{centering}
\end{figure}

 In Fig. \ref{fig:sec_prob}, we plot success probability using our codebook for uplink communication. We consider single transmit and receive antenna red and violet triangle marker are results obtained for $M=8$ and $M=4$ respectively after fine correction. We can obtain better results for same number of training symbols for $M=8$ with fine correction (red line) and $M=4$ without fine correction (blue line) as given in \cite{Zhang2020}.

\begin{figure}[h]
\vspace{1em}
\begin{centering}
\includegraphics[scale=0.5]{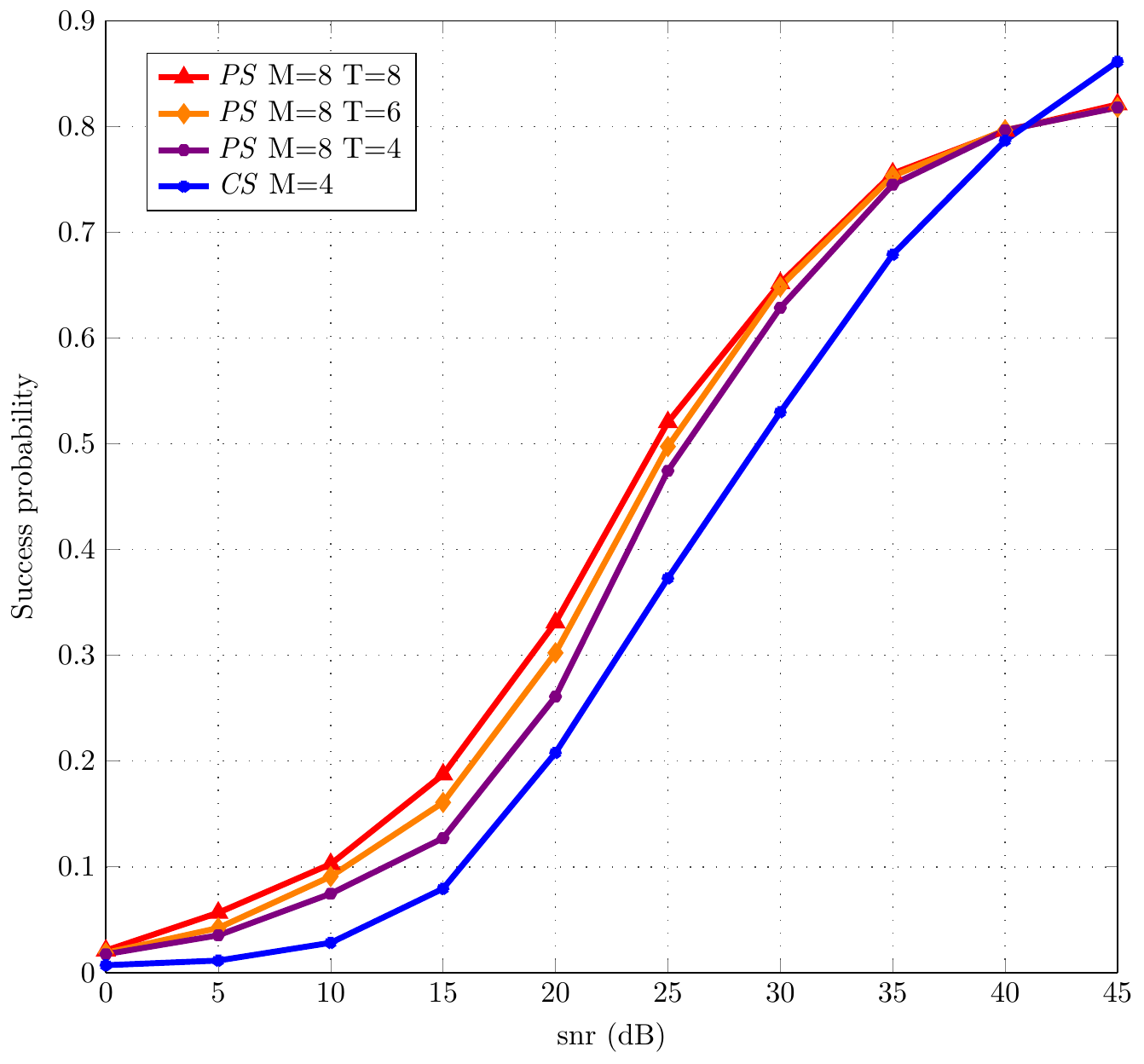}
\vspace{1.5em}
\caption{Success rate versus SNR for refinement over different number of bins in last round}
\label{fig:sec_prob1}
\end{centering}
\end{figure}

\indent\indent In Fig. \ref{fig:sec_prob1}, we plot success probability using our codebook for different values of $T$. Using $N_t=1$, $N_r=64$ and $N_h=160$, we obtain better results than codebook of \cite{Zhang2020} even with lesser training symbols by reducing the value of $T$. As in our codebook, the beam becomes narrower in successive rounds that reduces the chances for false detection and minimum intra bin distance does not reduces after round two. In the last round, the beam is very narrow and increases the probability of correct detection. In this method we are saving our training symbols for later search. By reducing the value of $T$ does not affect success probability at higher SNR as we are neglecting those bins which yields very low SNR. 

\begin{figure}[h]
\vspace{1em}
\begin{centering}
\includegraphics[scale=0.5]{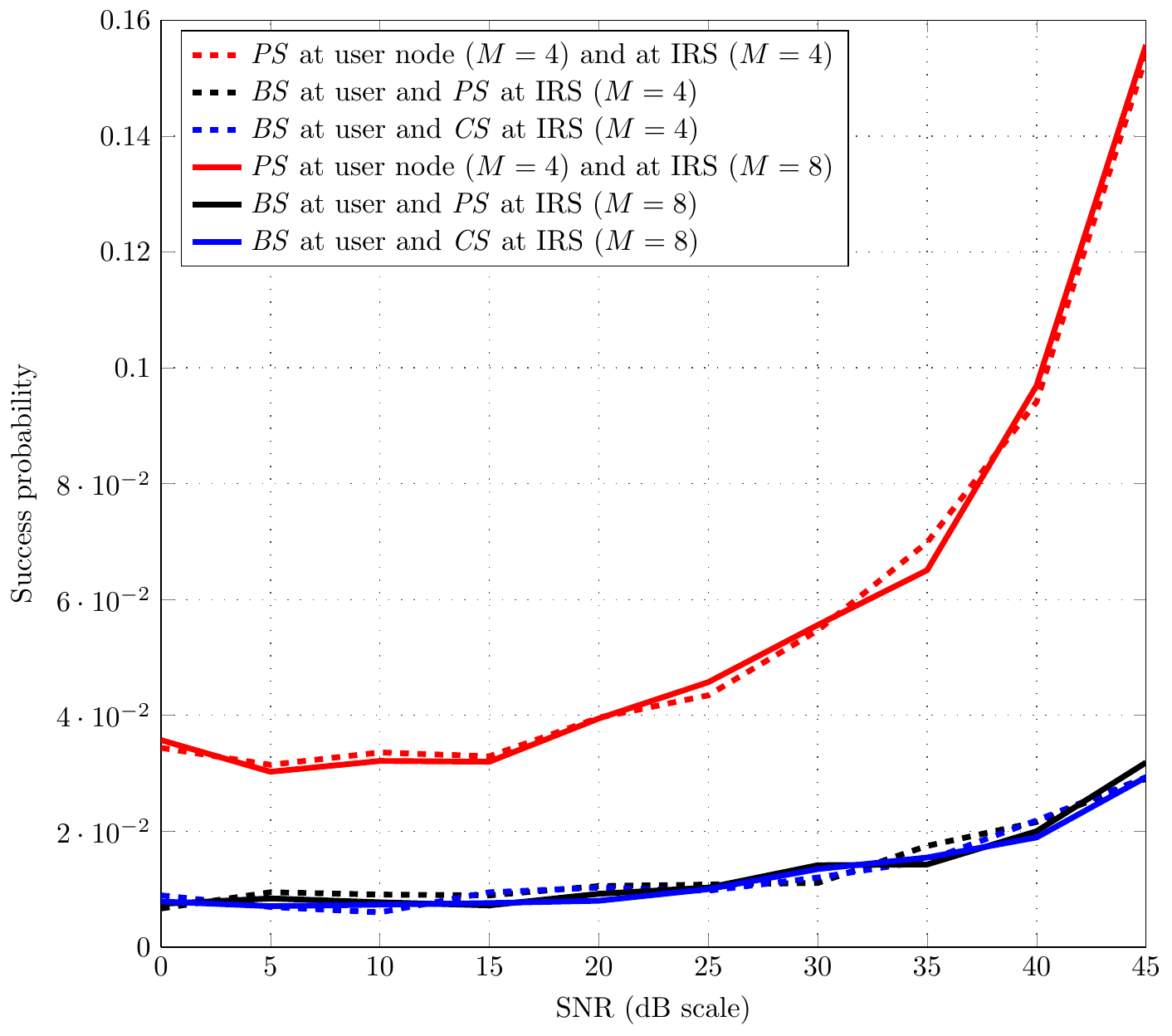}
\vspace{1.5em}
\caption{Success rate versus SNR for different schemes at user node}
\label{fig:sec_probuser}
\end{centering}
\end{figure}

 In Fig. \ref{fig:sec_probuser}, we apply our codebook for downlink communication after calibrating RIS for uplink communication. In this process we try to maximize the receive SNR at the user node by tuning antenna phase shifts at user node. Using $N_t=32$, $N_r=64$ and $N_h=160$, we obtain success probability using \emph{PS} and binary search (\emph{BS}) given in  \cite{Xiao2016}. It must be noted that success rate at the user node is very poor for RIS-assisted communication. Reason is that we are transmitting very low SNR and received SNR is getting better due to high gain from RIS while there is no significant improvement in SNR from received beamforming compared to passive beamforming at RIS. It shows that user node with a single antenna is equally good for RIS assisted communication.

\begin{figure}[h]
\vspace{1em}
\begin{centering}
\includegraphics[scale=0.5]{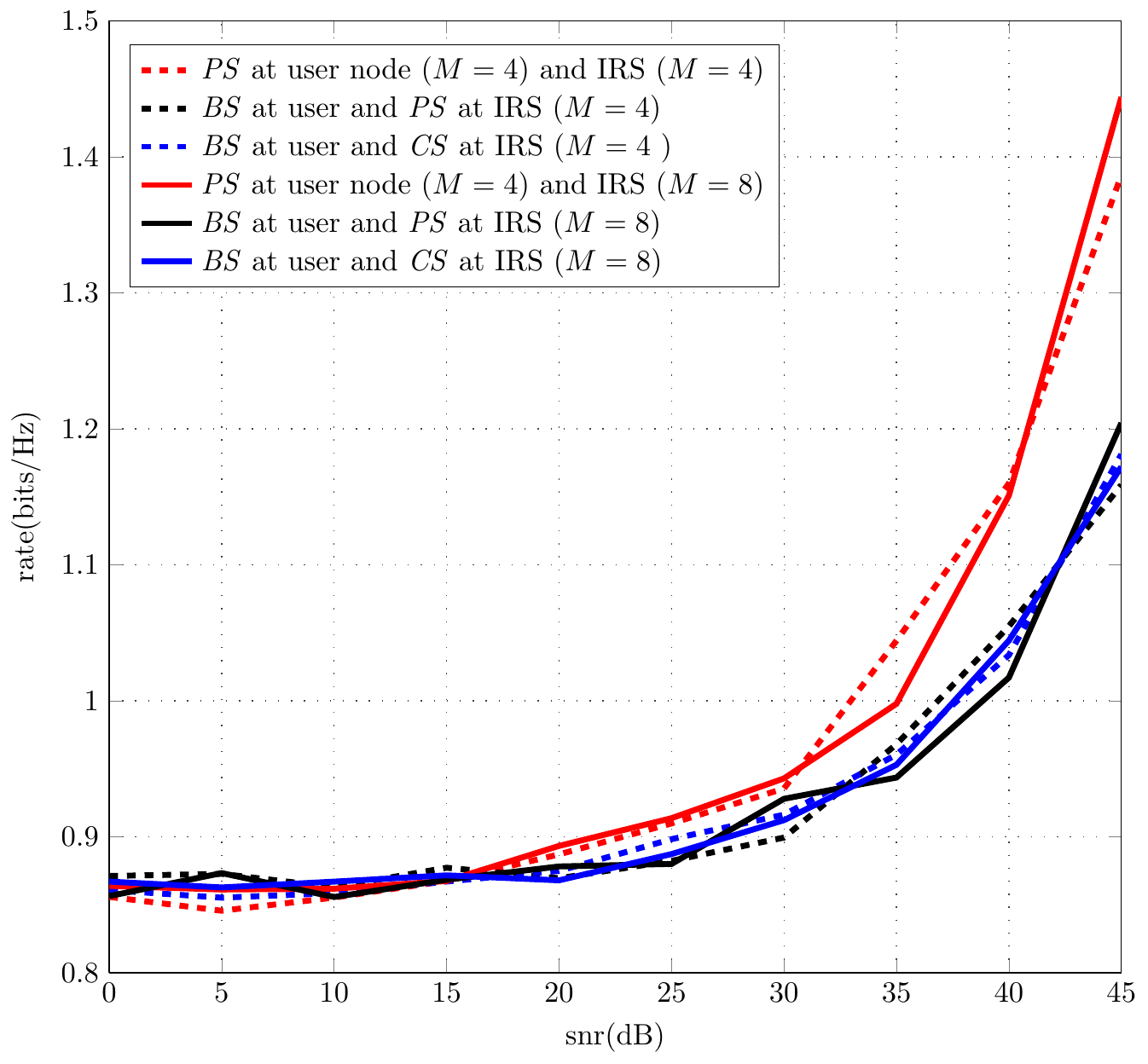}
\vspace{1.5em}
\caption{Rate versus SNR for different schemes at user node}
\label{fig:rate_curve}
\end{centering}
\end{figure}

 In Fig. \ref{fig:rate_curve}, we plot the rate after employing different beamforming schemes at the RIS and the user node. Using $N_t=32$, $N_r=64$ and $N_h=160$, we obtain the rate using our method and methods given in \cite{Zhang2020} \cite{Xiao2016}. The red line shows the rate obtained by employing our method at RIS and the user node. Black line is the rate obtained by using the proposed method at RIS while \emph{BS} at the user node given in \cite{Xiao2016}. Blue line shows results by employing codebook given in \cite{Zhang2020} at RIS and \emph{BS} for beam identification at user node. It shows that at low SNR results are almost similar by employing any of the codebook while as the SNR increases, fine correction in the last round of beam sweeping helps in beam identification.

\section{Conclusion}
\label{sec:con}
In this work, we have proposed the codebook for beam identification with refining the beam search for RIS assisted uplink communication. We find that refining the search in last round of codebook improves detection probability for the single user case. We observe that by increasing the number of beams in a single sweep, beam width increases and it increases the chance of wrong beam detection at the cost of reducing training symbols. But errors due to broadening of beam width can be corrected in last round with a narrow beam and hence increases the beam detection probability. Further we can optimize number of bins to be selected for later stage depending on size of RIS and number of beams used in a training symbol in the first round. For RIS with less number of reflecting elements, it is better to use low number of beams initially for sweep to maximize the detection probability. For large size RIS, number of beams are more in first round of beam sweep and we do fine correction in the later stage to maximize detection probability. 

RIS assisted uplink communication for multi user network has problems of beam collision and interference issues at AP. Hence multi beam codebook has to identify the collision event and AP has to inform the colliding nodes to stop transmission. In multi user case, if two users are belonging to different directions in same training symbol then AP might receive low SNR due to destructive interference and it does not identify the presence of users. Hence, codebook for uplink communication should be designed to identify interfering user nodes in multi user networks. 

\section{Acknowledgements}
\label{sec:ack}
Author also thanks the support from "NCKU 90 and Beyond" initiative at National Cheng Kung University, Taiwan.

\balance

\end{document}